\documentclass[sigconf]{acmart}

\copyrightyear{2022}
\acmYear{2022} 
\setcopyright{rightsretained}
\acmConference[ASE '22]{37th IEEE/ACM International Conference on
Automated Software Engineering}{October 10--14, 2022}{Rochester, MI,
USA}
\acmBooktitle{37th IEEE/ACM International Conference on Automated
Software Engineering (ASE '22), October 10--14, 2022, Rochester, MI,
USA}
\acmDOI{10.1145/3551349.3556926}
\acmISBN{978-1-4503-9475-8/22/10}

\usepackage{xspace}

\newcommand{\approach}{{\sc SelfAPR}\xspace}
\usepackage{newfloat,caption,subcaption,listings,xcolor,color,hyperref}
\usepackage{algorithm}
\usepackage{algorithmic}
\usepackage{amsmath}
\usepackage{numprint}
\usepackage{multirow}
\definecolor{light-gray}{gray}{0.97}

\newcommand{\CASE}[1]{\STATE \textbf{case} #1\textbf{:} \begin{ALC@g}}
\newcommand{\ENDCASE}{\end{ALC@g}}

\newcommand{\DEFAULT}{\STATE \textbf{default:} \begin{ALC@g}}
\newcommand{\ENDDEFAULT}{\end{ALC@g}}
\newcommand{\DEFAULTLINE}[1]{\STATE \textbf{default:} }

\usepackage[]{xcolor}
\newcommand{\TODO}[1]{\textcolor{red}{#1}\GenericWarning{}{LaTeX Warning: TODO: #1}}\newcommand\todo\TODO

\usepackage{pifont}

\definecolor{bluekeywords}{rgb}{0.13,0.13,1}
\definecolor{greencomments}{rgb}{0,0.55,0.2}
\definecolor{redstrings}{rgb}{0.9,0,0}

\newcommand{\ASECMR}[1]{\textcolor{black}{#1}}

\DeclareFloatingEnvironment[fileext=frm,placement={!ht},name=Listing]{listing}
\begin{document}

\title{SelfAPR: Self-supervised Program Repair  \\with Test Execution Diagnostics}

\author{He Ye}
\email{heye@kth.se}
\affiliation{
  \institution{KTH Royal Institute of Technology}
  \country{Sweden}
}

\author{Matias Martinez}
\email{matias.martinez@uphf.fr}
\affiliation{
  \institution{Université Polytechnique Hauts-de-France}
  \country{France}
}

\author{Xiapu Luo}
\email{csxluo@comp.polyu.edu.hk}
\affiliation{
  \institution{The Hong Kong Polytechnic University}
  \country{China}
}

\author{Tao Zhang}
\email{tazhang@must.edu.mo}
\affiliation{
  \institution{Macau University of Science and Technology}
  \country{China}
}

\author{Martin Monperrus}
\email{monperrus@kth.se}
\affiliation{
  \institution{KTH Royal Institute of Technology}
  \country{Sweden}
}

\begin{abstract}
Learning-based program repair has achieved good results in a recent series of papers.
Yet, we observe that the related work fails to repair some bugs because of a lack of knowledge about 1) the application domain of the program being repaired, and 2) the fault type being repaired.
In this paper, we solve both problems by changing the learning paradigm from supervised training to self-supervised training in an approach called \approach.
First, \approach generates training samples on disk  by perturbing a previous version of the program being repaired, enforcing the neural model to capture project-specific knowledge. This is different from the previous work based on mined past commits.
Second, \approach executes all training samples and extracts and encodes test execution diagnostics into the input representation, steering the neural model to fix the kind of fault. This is different from the existing studies that only consider static source code as input. 
We implement \approach and evaluate it in a systematic manner. We generate \numprint{1039873} training samples obtained by perturbing 17 open-source projects. We evaluate \approach on 818 bugs from Defects4J, \approach correctly repairs 110 of them,
outperforming all the supervised learning repair approaches. 
\end{abstract}

% \keywords{self-supervised learning, automated program repair}

\maketitle

\section{Introduction}
\label{sec-introduction}

Automated program repair (APR) aims to reduce the manual and costly work related to the bug localization and bug fixing tasks of software maintenance~\cite{Monperrus2015,TSE-repair-survey,nollertrust-icse22}. While early works in the field mainly used
search-based \cite{LeGoues2012GenProg,tbar} or semantics-based \cite{Angelixicse16,concolic-repair-PLDI21} techniques, recently, a different line APR research has proven successful: neural machine translation for program repair, or simply ``neural program repair'' \cite{codit-tse20,SEQUENCER,RewardRepair-icse22,Tufano-ICSE19,modit-ase21,CoCoNuT,Recoder}.

Neural program repair is based on a common encoder-decoder architecture to  transform the buggy code to the correct code, yet the proposed approaches differ as follows:
1) In the input or output representations, for example, the  decoders may output code edits \cite{Recoder};
2) In the pre-processing or post-processing phases of the data, for example filtering the patches with invalid identifiers \cite{CURE-icse21}.
When those past works are compared one against the others on the same benchmark \cite{RewardRepair-icse22}, those variations explain the performance differences.

% supervised learning
Despite those differences, the previous work on neural program repair is dominantly founded on the same machine learning paradigm: supervised learning \cite{Hinton:2007}.
In that context, the supervised training samples come from mining real-world commits made by human developers. 
For example, Recoder's \cite{Recoder} training samples were downloaded from GitHub, totaling \numprint{1083185} commits done between March 2011 and March 2018. 
In this paper, we claim and provide evidence that this supervised paradigm for neural program repair has two fundamental limitations as follows.

% https://drive.google.com/file/d/1pcx-FKO_RL5CdNY67_PI-ozye51ixeeB/view?usp=sharing
\begin{figure}[t!]
\includegraphics[width=0.338\textwidth,height=2.38cm]{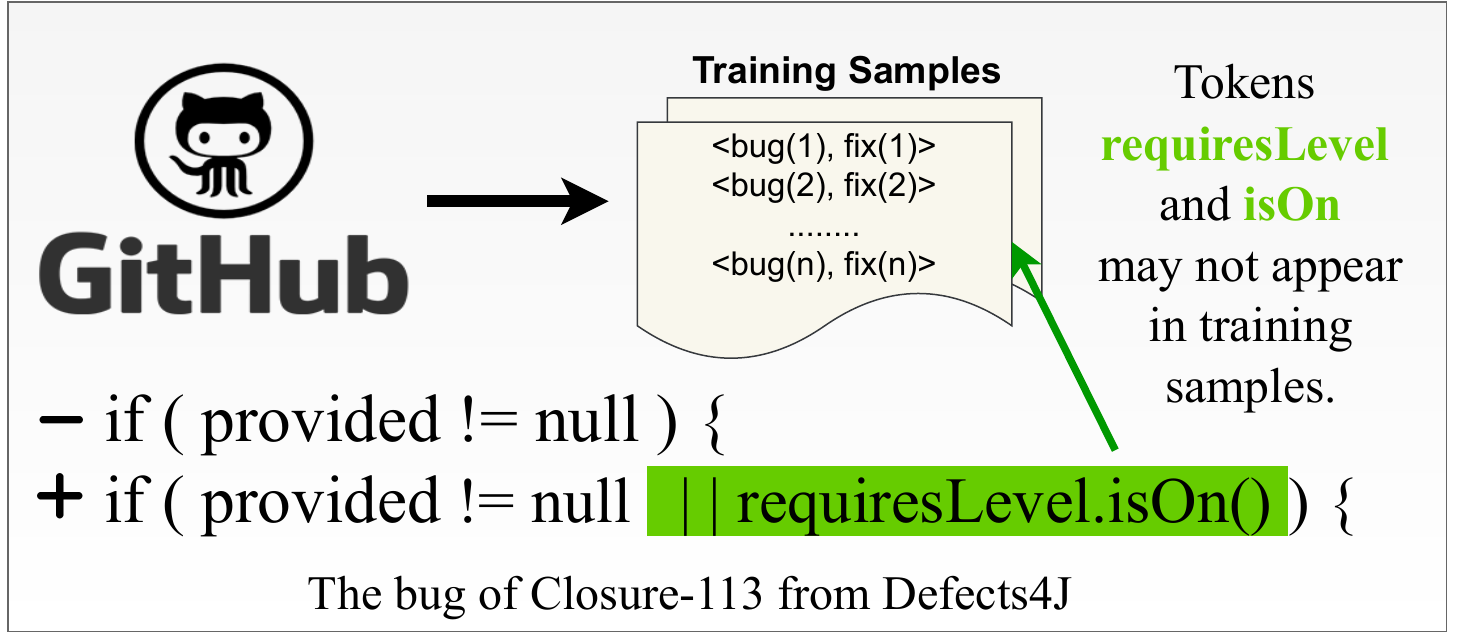} 
 \caption{Motivating example of \approach: supervised neural program repair is fundamentally limited by the absence of project-specific tokens at training time.}
\label{fig:motivating}
\end{figure}

% https://drive.google.com/file/d/1AAqXESQsZF6ehST2xZo4_3cw8cjctoZ6/view?usp=sharing
\begin{figure*}[t!]
\includegraphics[width=0.75\textwidth,height=0.228\textheight]{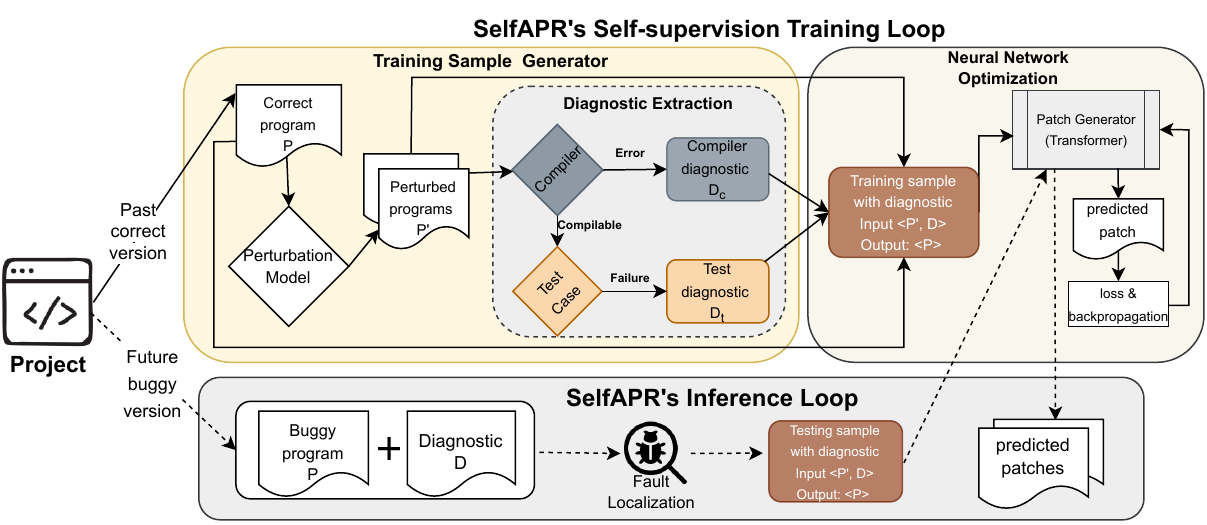} 
 \caption{Overview of \approach: the key novel features are the training sample generator and the diagnostic in the input representation. }
\label{fig:overview}
\end{figure*}
 
\textbf{Problem 1: Lack of project-specific knowledge.} Past commits used for training are typically collected from open-source projects. 
Those projects may have little or nothing to do with the application domain of the program under repair. This is  known as the exposure bias problem \cite{exposure-bias-Bengio-NIPS15}. 
In other terms, the training samples do not contain project-specific knowledge. By project-specific knowledge, we refer to specific fixing tokens of variables, expressions and statements, and their semantic relationships that the neural network can use to synthesize a patch. 
\autoref{fig:motivating} illustrates this point. Bug \texttt{Closure-113} from Defects4J benchmark \cite{defects4j} is fixed by expanding a boolean expression with  \texttt{requiresLevel.isOn()}. There are few projects in the wild that use such tokens. Indeed, none of the existing work on neural program repair can fix bug  \texttt{Closure-113}, because the project-specific  tokens \texttt{requiresLevel} and \texttt{isOn} have not been seen in training samples. Note that this is the core conceptual limitation, not a quantitative one: crawling more training samples from Github would not solve this problem.

\textbf{Problem 2: Lack of execution information}.
Recently,  Chakraborty and Ray \cite{modit-ase21} have shown that neural repair models fail because of a lack of guiding information about the bug type. Our key insight is that execution information is rich in such signals. 
For instance, it may be useful for the network to know that the bug to be repaired is a \texttt{NullPointerException}.
For supervised training with past commits, it is virtually impossible to obtain this information at scale because: 1) it is very hard to compile past commits due to the version and dependency hell \cite{dependencyhell}; 2) many past commits do not come with a way to execute the bug, such as a test suite.

To address the above two problems, we devise, implement and evaluate \approach. 
The radical novelty of \approach is to be based on self-supervised learning instead of supervised learning.
\approach consists of an original training loop on a past  version of the project under repair.
The core intuition is that the past version contains a wealth of useful project-specific knowledge.
The self-supervised learning loop means generating training samples ourselves (instead of mining commits), by perturbing an old version of the project under repair in a careful and systematic manner. This is a potent solution to solve our two problems.

%\textbf{\emph{Self-old} version enables project-specific knowledge to be learned.} 
\textbf{Self-supervised training enables us to learn project-specific knowledge.} 
\approach leans from a past version of the program to be repaired. 
From a single past version, \approach generates thousands of synthetic training samples, which all contain project-specific tokens by construction. 
This allows \approach to learn project-specific knowledge, including rare tokens, idiomatic expressions, and semantic relationships between domain identifiers (types, variables, and methods).
As we shall see later, the bug \texttt{Closure-113} shown in \autoref{fig:motivating} is correctly fixed by \approach, because \approach learns to use project-specific tokens from the self-supervised training samples (recall that no previous work has succeeded in repairing this bug).

\textbf{Self-supervised learning enables us to add the execution information into training samples.}
By generating thousands of samples from a single, working, executable version of the project, it becomes doable to compile and test all training samples. 
In other words, the powerful concept of self-supervision opens the door to embedding execution information in the training loop.
Consequently, \approach proposes a novel input representation that includes the execution information, called ``diagnostic'' for short in the rest of this paper. Put simply, \approach's self-supervised paradigm augments the training samples from \texttt{<bug, fix>} to \texttt{<bug, execution diagnostic, fix>}.

%what we do
Overall, \approach works as follows. It first generates thousands of training samples based on perturbation.
Next, each perturbed program is executed to see whether it is buggy (does not compile or does not pass the test). 
For each buggy program, we extract the diagnostic and incorporate it into the input representation. Next, we train \approach with a state-of-the-art transformed-based neural model \cite{Vaswani2021Transformer} with this input, using the correct program before perturbation as the expected output.
At inference time, we use the error of the failing test case of the program under repair as input diagnostic. 

% evaluation
We evaluate our work on 818 bugs from 17 open-source projects from Defects4J \cite{defects4j}. 
In total, we generate \numprint{1039873} training samples in a self-supervised manner, each of which contains an error diagnostic.
% result
Our experimental results show  that
\approach succeeds in repairing 65/388 bugs from Defects4J version 1.2 and 45/430 bugs from  Defects4J version 2.0, which is a clear improvement over the state-of-the-art.  \approach is able to repair 10 bugs that have never been reported to be repaired by the related supervised learning repair approaches.
More importantly, generating perturbation-based training samples based on a past version of the project under repair contributes 30\% effectiveness of \approach, thanks to \approach's unique capability to learn and reuse project-specific tokens in the synthesized patch. 
In other terms, this performance breakthrough is due to the paradigm shift from supervised neural networks to self-supervised neural networks.

To sum up, we make the following contributions: 
\begin{itemize}
% concepts
\item  We devise, \approach, an original self-supervised neural model for program repair based on execution. 
To our knowledge, this is the first learning architecture that encodes and leverages test execution diagnostics for repair.
Notably, \approach is able to learn project-specific repair knowledge in an effective way.

% experiments
\item We perform an original series of experiments and show that  \approach repairs 110 bugs from 17 open-source projects.
Our experimental results demonstrate that self-supervised learning on a past version of the program under repair significantly increases the repair effectiveness.  
Our experiments show that including project-specific training samples directly contributes to repairing 20 more bugs (+30\%) on Defects4J version 1.2.

% code and data
\item We consolidate and share our training dataset of \numprint{1039873} buggy training samples, including \numprint{408858} functional errors with at least one failing test case and \numprint{631015} compilation errors. 
This dataset is valuable for future researchers to understand the syntactic and semantic relationships between errors and code changes responsible for them.  
We make all our code and data publicly available at \color{blue}{\url{https://github.com/SophieHYe/SelfAPR}}.

\end{itemize}

\section{Background}
\label{sec:background}

\subsection{Automated Program Repair (APR)}

\textit{Search-based Repair.} Search-based approaches  such as GenProg \cite{LeGoues2012GenProg}, SPR \cite{SPR-FSE15} and others \cite{Yuan2017ARJAAR,astor,tbar,hercules,varfix,quixbugs-jss}, typically generate a large number of tentative patches according to different edit patterns, such as inserting null checks \cite{NPEFix},  mutating operators \cite{OperatorMutation} or copying statement \cite{LeGoues2012GenProg}.
In the patch search space, they then employ heuristic search algorithms (e.g., genetic programming)  to quickly find patches that can pass the given test specification. 
Search-based approaches suffer from both the immense scale of search space and the effectiveness of search algorithms \cite{QiIssta15-overfitting}.

% Semantics-based Repair
\textit{Semantics-based Repair. } Semantics-based  approaches, such as Angelix \cite{Angelixicse16}, S3 \cite{s3} and others \cite{acs,nopol,directfix,CrashProgramRepair-ISSTA19,concolic-repair-PLDI21,patch-transplantation-TOSEM21,sergey-test-equivalence-tosem18,semfix},  first construct a constraint problem 
that should be satisfied to fix a bug, and then they use some kind of program synthesis to synthesize patches that satisfy the repair constraints. Semantics-based approaches effectively narrow down search spaces, yet they mostly limit the edit patterns to small-scale expressions in order to make program synthesis tractable \cite{varfix}.

\textit{Learning-based Repair.} Learning-based approaches based on supervised learning \cite{genesis,compilation-error-fse19,SEQUENCER,CoCoNuT,Tufano-ICSE19,codit-tse20,Tufano-tse19,Recoder,CURE-icse21,modit-ase21,RewardRepair-icse22,codexrepair-nus,tfix} are data-driven, employing pairs of the buggy and fixed code crawled from open-source projects and learning the edit patterns from a large scale training dataset. 
Learning-based approaches typically rank the patches based on the maximum likelihood estimation of tokens, which potentially narrows down the search space to likely patches. Nevertheless, learning-based approaches treat source code as natural language translation, suffering a lack of knowledge in programming languages.

\subsection{Self-supervised Learning}

% novelty claim, may be to be moved a little after
%Compare to other self-supervised learning

Self-supervised learning is the idea to transform unlabeled  data into labeled data, without any human labeling.  Self-supervised learning addresses the key limitations of supervised learning when it comes to collecting, handling, cleaning, and labeling training data \cite{word-representation}.
In NLP, self-supervised learning has  been used with great success for learning word representations \cite{glove}.
In machine learning on code, it is powerful to capture contextual representations \cite{NIPS2015_7137debd,acl-18-language-model,codebert,plbart} and the typical perturbation strategies are masking out or replacing tokens.

To our knowledge, the idea of self-supervised learning in program repair is largely unexplored. Only Yasunaga and Liang \cite{Yasunaga20DrRepair} and Allamanis et al. \cite{allamanis2021self-buglab} have done preliminary investigations, which will be discussed in Section \ref{sec:rw}.

\section{\approach}
\label{sec-approach}

\begin{figure}[t!]
\includegraphics[width=0.438\textwidth,height=6.6cm]{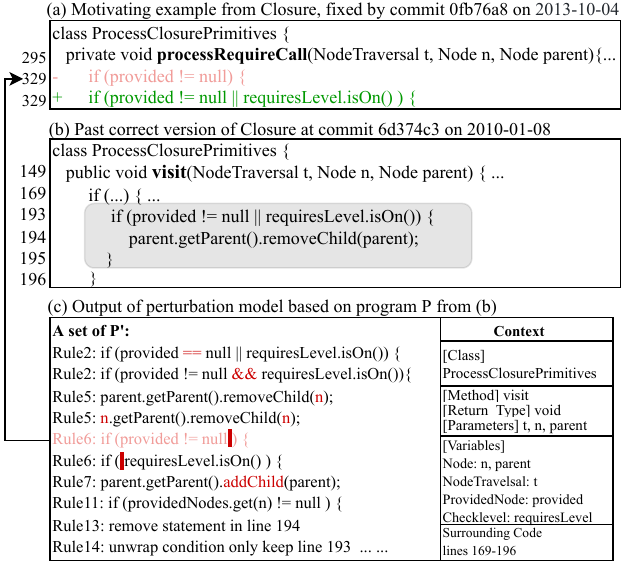} 
 \caption{A running example of the perturbation model.}
 
\label{fig:running-example}
\end{figure}

\subsection{Overview}
\label{sec-overview}
% \approach in one paragraph

\autoref{fig:overview} gives an overview of \approach, where the upper part shows the training phase and the bottom part presents the inference phase.
The core idea of \approach is to generate training samples in a fully project-specific and self-supervised manner.
In order to be capable of doing project-specific training, we take one  project's past version  to generate training samples. The learned model is used to repair  future bugs from the same project. 
For self-supervised training, the training buggy samples are generated by perturbing correct programs given as input.
This is contrary to the related work on neural program repair which is based on supervised learning with mined past commits \cite{Tufano-ICSE19,Tufano-tse19,SEQUENCER,CoCoNuT,CURE-icse21,Recoder,RewardRepair-icse22}. Notably, the perturbation-based programs can be compiled and executed with project-specific test suites while past commits do not.

\emph{Training Phase.} 
\approach's training phase consists of two main components: a training sample generator based on perturbing the source code (left part) and a neural transformer architecture fed with an input representation embedding  test execution information (right part).
%main loop
% core input / assumption of the technique
Given a correct program $\mathcal{P}$ which can be compiled and executed (i.e., pass test suite specification), \approach generates variants of them, called here ``perturbation-based programs'' (denoted as $\mathcal{P'}$), according to a perturbation model (Section \ref{sec-perturbation}).
For one single program $\mathcal{P}$, a number of $\mathcal{P'}$ can be generated in a  configurable way, depending on the required size of the training dataset. All the generated $\mathcal{P'}$ are compiled and executed by invoking the test suite included in the correct version. Compilation and test execution determine  whether the perturbation model introduces a bug, and yields the error type and the error diagnostic $\mathcal{D}$.

Finally, \approach's learning model takes input as  $\mathcal{P'}$ and its accompanying diagnostic $\mathcal{D}$.
The goal of the machine learning model is to predict the expected output $\mathcal{P}$, the original source code before perturbation.  
They are denoted as follows:
$$
\scriptsize
training\_samples = 
\{ input: <\mathcal{P'}, \mathcal{D}>
   output: <\mathcal{P}>
 \}
$$

%briefly introduce each
\emph{Perturbation Model.} 
The goal of the perturbation model is to generate training samples with bugs.
% , where a sample is pair of programs <$\mathcal{P'}$, $\mathcal{P}$>, the perturbation-based program $\mathcal{P'}$ being synthesized by perturbing the correct program source code $\mathcal{P}$.  
% diagnostic
Perturbation-based programs are valuable for training a neural repair network if they can be considered as buggy in some sense.
To determine this, \approach employs a compiler and the available test suite to identify their correctness. 
If a perturbation-based program $\mathcal{P'}$ produces either a compiler or a test execution error, it is deemed as buggy, and consequently, used as a training sample later. The perturbation model will be described in Section \ref{sec-perturbation}.

% paragraph about diagnostics
% \todo{what problem does diagnostic collection solve?}
\emph{Diagnostic Extraction}.
Not only \approach uses the compiler and test execution to select buggy training samples, it also uses them to extract diagnostics $\mathcal{D}$ about the bug. 
The $\mathcal{D}$ is then a first-class part of the neural network input.
% intuition
Our intuition is that error diagnostics could be useful to guide patch generation, for example, the fact that a \texttt{NullPointerException}  is thrown provides explicit information regarding the repair action to be used. 
We note that this radically departs from the related work which only uses source code as input \cite{Tufano-ICSE19,Tufano-tse19,Recoder,CURE-icse21,RewardRepair-icse22}.

% paragraph about learning
\emph{Input Representation.}
A training sample is represented as a sequence of tokens.
%context
We follow the existing neural program repair \cite{modit-ase21,CURE-icse21,CoCoNuT,RewardRepair-icse22} to include the context code of $\mathcal{P'}$ in the input representation. 
The context code is enriched with a summary of the following information: 1) class and method name, 2) variables in the buggy method scope and 3) the surrounding source code. 
%diagnostic
For diagnostic $\mathcal{D}$, we concatenate the input representation of diagnostics as a sequence of tokens, which is coming from the compiler or test suite execution failures.

\emph{Subtokenization.}
Once a training sample is represented as a sequence of tokens, we use subtokenization before entering it into the neural network.
This is essential in order to reduce  vocabulary size \cite{CURE-icse21}.
\approach follows \cite{modit-ase21,RewardRepair-icse22} to use a sentence-piece tokenizer \cite{Kudo2018SentencePieceAS}.
Sentence-piece tokenization divides every token into a sequence of subtokens.
% For example, the token \texttt{GET\_FACTORS} would be subtokenized as three words \texttt{GET}, \texttt{\_}, and \texttt{FACTORS}.

\emph{Inference Phase.}
In the inference phase, we use the trained model to repair future bugs from the same project.
Specifically, \approach takes an input of a buggy program ($\mathcal{P}$) and the failing test diagnostic $\mathcal{D}$. 
We employ fault localization (FL) (e.g., Ochiai \cite{fl-tool} or Gzoltar \cite{GZoltar}) to generate a ranked list of suspicious buggy lines. 
For a given suspicious statement found by FL,  \approach constructs an inference input per our representation with: 1) the suspicious statement and its context, and 2) the test execution diagnostic.
This input is given to the patch generator (transformer-based neural model), which outputs the most likely patch, or may enumerate the $K$ best patches for that suspicious statement with beam search, where  $K$ is fully configurable, a.k.a, the beam search size.

% novelty claim

To our knowledge, our work is  novel in two aspects.
First, it is the first to propose project-specific training using a different version of the same project under repair. This enables the neural model to learn project-specific knowledge and mitigate the training discrepancy when the training and testing datasets come from different projects.
Second, \approach is the first to add compiler and test suite execution information as part of the input representation for neural program repair. 
As an opposite,  the previous supervised-learning studies all consider as input a pair of the buggy and fixed source code.
The related two self-supervised learning approaches \cite{Yasunaga20DrRepair, allamanis2021self-buglab} neither  include test execution diagnostics, nor are founded in project-specific training with a past version of the project under repair.

% % https://drive.google.com/file/d/16AGDv-mG3zqqoSuzeVql_dWWd6aGCIHU/view?usp=sharing
% \begin{figure}[t!]
% \includegraphics[width=0.48\textwidth]{Figures/perturb.drawio.pdf} 
%  \caption{The workflow of perturbation model.}
% \label{fig:perturb}
% \end{figure}

\subsection{Perturbation Model}
\label{sec-perturbation}
%WHY

The goal of the perturbation model is to generate training samples with bugs. 
Recall that the perturbation model generates bugs from a correct  program. We train the neural model with the buggy programs as input and learn to generate the correct version as output.
In \approach, we design the perturbation rules ($Rules$) driven by learning objectives that we expect the neural model to learn. For example, if the learning objective is to learn how to insert statements, a perturbation model may generate training samples by deleting statements: reversed, the training sample teaches the neural model to learn inserting code that has been deleted.

\subsubsection{Running Example}

\autoref{fig:running-example} gives a running example to demonstrate how the perturbation model creates the perturbation-based programs ($\mathcal{P'}$) and the corresponding context information. 
Recall that the motivating example (\autoref{fig:motivating}) from project Closure was fixed by the developer on 2013-10-04, shown in \autoref{fig:running-example}(a), by adding a clause to an if condition.
To create training samples that are useful for this bug, \approach perturbs a past version of Closure and removes a clause in some selected if conditions.
We note that although the buggy method \texttt{processRequireCall} in (a) does not exist in the past version, there are similar methods that could be perturbed in order to create training samples. 

\autoref{fig:running-example}(c) shows the output of \approach's perturbation model with a set of $\mathcal{P'}$ at locations 193 and 194 of (b), as well as the context information including class, method, variables and surrounding code. 
The set of $\mathcal{P'}$ are generated based on different perturbation rules that are described later. 
One sample generated using $Rule_{6}$ is a training sample that is useful to learn to fix the bug in \autoref{fig:motivating}.

From this running example, we see that first, for one single AST statement, multiple perturbation-based programs are generated. 
Second, by learning on a past correct version of the project under repair, \approach creates valuable training samples encoded project-specific knowledge to fix a new and unseen bug appearing three years later.

\begin{listing}[h!]

\begin{lstlisting} [firstnumber=419] 
<@{\underline{$Rule_{1}$:} modify declaring type ...}@>
    - double a;
    + float a;
<@{\underline{$Rule_{2}$:} modify operator ==, !=, \&\&, ||, +, -,*,\%, ...}@>
    - if( a > b )
    + if( a >= b )
<@{\underline{$Rule_{3}$:} modify literal, "STRING", true, false, 1, 0,...}@>
    - return a.length-2;
    + return a.length-1;
<@{\underline{$Rule_{4}$:} modify constructor}@>
    - new ClassA(a,b)
    + new ClassB(b,c)
<@{\underline{$Rule_{5-1}$:} modify argument}@>
    - invoc1(a, b)
    + invoc1(a, x)
<@{\underline{$Rule_{5-2}$:} swap argumens}@>
    - invoc1(a, b)
    + invoc1(b, a)   
<@{\underline{$Rule_{6-1}$:} reduce boolean expression}@>
    - if (exp1 && exp2 )
    + if (exp1)
<@{\underline{$Rule_{6-2}$:} expand boolean expression}@>
    - if (exp1 && exp2 )
    + if (exp1 && exp2 || exp3)
<@{\underline{$Rule_{7-1}$:} modify invocation}@>
    - a.invoc1(a, b)
    + a.invoc1(a, b, c)
<@{\underline{$Rule_{7-2}$:} replace invocation}@>
    - a.invoc1(a, b)
    + a.invoc2(a)
<@{\underline{$Rule_{8}$:}  compound of $Rule_{1}$ - $Rule_{7}$  }@>
    -  if (exp1 > 1 && exp2 )
    +  if (exp1 >= 0 || exp2 == null )
<@{\underline{$Rule_{9}$:} replace by transplanting a similar donor statement}@>
    - target statement
    + donor statement 
<@{\underline{$Rule_{10}$:} move a later statement before the target statement}@>
    +  later statement;
        target statement;
    -  later statement; 
<@{\underline{$Rule_{11}$:} transplanting a donor statement}@>
        target statement;
    +  donor statement;
<@{\underline{$Rule_{12}$:} wrap target statement with an existing conditional block}@>
    + if (exp1) {
            statement1;...
    +}
<@{\underline{$Rule_{13}$:} insert an existing block (if, loop, etc)}@>
    + if(exp1) {
    +   statement1;...
    + } 
<@{\underline{$Rule_{14}$:} delete statement}@>
        if (exp1!=null && exp1>exp2) {
    -   statement 1; ...
        }       
<@{\underline{$Rule_{15}$:} unwrap block}@>
    - if (exp1!=null && exp1>exp2) {
            statement1; statement2;
    - }     
<@{\underline{$Rule_{16}$:} remove block}@>
    - for (exp1) {
    -     statement; ...
    - }
\end{lstlisting} 
\caption{Perturbation rules for self-supervised training.}
\label{listing-replace}

\end{listing}

			 %//1.replace executable
			 
			 %//2.replace arguments.
			 
			 %//3.replace with overriding executable

\subsubsection{Perturbation Rules}
In \approach, we systematically design perturbation rules (called as $Rule$ in the following) according to learning objectives.
\autoref{listing-replace} lists the perturbation rules and we now explain them in detail.

% Higher level training objectives:

% Learning the scoping rules
% Learning the project-specific typing rules
% Learning to prioritize
% Learning to synthesize code (=insert)
% Learning to remove code
% Learning to reuse existing code

\ASECMR{
\textbf{Learning objective: learning to only use valid identifiers.}
The first learning objective is enabling the neural network to understand how to use valid identifiers according to scoping and typing rules. The following rules are designed accordingly.
}

\ASECMR{
$Rule_{1}$ perturbs a correct declaring type with a wrong one. 
\approach takes specific care of the interchangeability between the types according to the type hierarchy (e.g., replacing Set with List). 
$Rule_{2}$ perturbs the correct  operator with the wrong one. If an AST statement has more than one operator, \approach iterates over each of them. 
$Rule_{3}$ perturbs the correct literal with the wrong one. 
The literal replacement follows typing constraints.
$Rule_{4}$ perturbs the correct constructor with the wrong or an overloading one, where the added constructor has already been used in the same class file. \approach chooses the wrong constructor with the required variables following typing and scoping constraints.  
 $Rule_{5}$ perturbs variables:
 $Rule_{5-1}$ modifies a correct variable with a wrong one, following typing and scoping constraints. Moreover, $Rule_{5-2}$ swaps two arguments if they share the same type.
$Rule_{7}$ perturbs an invocation:
$Rule_{7-1}$ modifies the correct invocation with an overloading one (if there exists one). 
$Rule_{7-2}$ replaces the correct invocation with a new one that appears in the class file. For both rules, \approach synthesizes the arguments following the typing and scoping constraints of available variables.
$Rule_{8}$ generates a compound statement by stacking $Rule_{1}$ - $Rule_{7}$ from a correct statement in order to increase the complexity of the learning task for some training samples.
}

\ASECMR{
\textbf{Learning objective: learning to reuse existing code from the same program.}
Ever since GenProg \cite{LeGoues2012GenProg}, it is known that reusing code from the program repair is valuable \cite{1403.6322,BarrBDHS14}. Hence, we want perturbations that encourage the neural network to reuse code from in the close vicinity of the buggy location. 
}

\ASECMR{
$Rule_{6}$ are dedicated to boolean expressions. $Rule_{6-1}$ perturbs boolean expressions by removing a clause. 
$Rule_{6-2}$ expands a correct boolean expression by transplanting an existing binary expression in the method scope, encouraging clause reuse. 
$Rule_{9}$ modifies the correct statement by transplanting a similar statement taken from the class scope, called the donor statement.  In \approach, the donor statements are selected based on edit distance with the target statement,  and statements with a higher similarity score than the default threshold are selected. 
Notably, code transplantation is a more generic strategy than the others, it augments the diversity of the generated training samples behind the transformations by the previous $Rules$.
}

\ASECMR{
\textbf{Learning objective: learning to synthesize code according to the context.}
It is known that code has low entropy, because of its high contextual redundancy \cite{hindle2012naturalness}. We design the following perturbations in order to for the neural network to learn to synthesize a patch according to the closest surrounding code.
}

\ASECMR{
$Rule_{10}$ modifies the order of correct statements. It shuffles the target statement with one from the surrounding 3 context lines of the target statement.  
$Rule_{14}$ deletes the target statement. The learned repair actions are diverse depending on the characteristics of the deleted statement.
$Rule_{15}$ unwraps the condition and only keeps the  \texttt{then} branch. This enables the neural model to learn to generate conditions that are semantically related to the target statement to be wrapped.
$Rule_{16}$ deletes a complete AST block.  This enables the neural model to learn to generate complex and multi-line code.
}

\ASECMR{
\textbf{Learning objective: learning to delete.}
Deleting code is an option for repairing bugs \cite{QiIssta15-overfitting,ginelli2022}.
To train the neural network to delete code, we design perturbations that add superfluous code. Recall that the perturbed code is then given as input, meaning that the expected output is indeed a code removal.
}

\ASECMR{
$Rule_{11}$ inserts donor statements randomly before or after the statement under perturbation, where the donor comes from the surrounding code.
$Rule_{12}$ transplants a donor conditional expression and wrap a target statement. The conditional expression is taken from the class scope and with a textual similarity with the target statement.
$Rule_{13}$ transplants an entire code block before or after the target statement under perturbation. The block must exist in the class scope and with a default textual similarity with the target statement.
}

\emph{Sanity check of perturbation-based training samples.}
All training samples are validated with the following sanity check: 1) they are different from the correct program; 2) they are unique, i.e., we deduplicate the training samples even if different $Rules$ generate the same training samples; 3) they are buggy. 
Then, we guarantee that the perturbed code indeed triggers a bug by executing them against the compiler and test suite. 

\emph{Diversity of perturbation-based training samples.}
Our perturbation rules mix fixed transformations and generic ones. As a result, the generated training samples are diverse, we will give quantitative evidence in \autoref{fig:rule-distribution}. This diversity enables the network to learn a variety of repair actions that can go beyond the fixed transformations, as demonstrated in \autoref{fig:correlation}.

% ========================================================================
% ========================================================================
% ========================================================================
% ========================================================================
% ========================================================================
% ========================================================================

\begin{table*}[t!]
\footnotesize
\renewcommand{\arraystretch}{1}
%  \begin{tabular}{p{0.12\linewidth}|p{0.1\linewidth}p{0.07\linewidth}p{0.05\linewidth}lp{0.05\linewidth}p{0.05\linewidth}p{0.1\linewidth}|p{0.08\linewidth}r}

 \begin{tabular}{l|ccccccc|lr}

\hline
\multirow{2}{*}{Projects}& \multicolumn{7}{c}{Self-supervised Training} & \multicolumn{2}{c}{Testing on Real-word Bugs}\\

 & CommitID  &  Date  & LOC  & \# Test Func   & \# CE&\#  FE&\# Training Samples  & Date since & \# Bugs \\
\hline
Closure&6d374c3&2010-01-08&60875&5280 &142420 &91824 & 234244 &2010-02-05&173 \\

Chart&68e4916&2007-07-06&78566&2444     &127391&98562& 225953    &2007-08-28&25 \\
JacksonDatabind&88f44d8&2013-05-17&42965&2693  &96775&36946&  133721 &2014-05-28&111\\
Time&e0559c5&2010-12-05&26795&5012    &68910&39875& 108785  &2011-02-15&25 \\
Lang&bb16716&2006-07-21&16623&2522   &44912 &24098  & 69010  &2006-08-18&63 \\
JacksonCore&b40ac81&2013-08-28&15882&354   &37198 & 23836 &  61034 &2013-09-21&25 \\
JxPath&fab38ab&2007-01-10&19373&441   &24741  &22325   & 47066  &2007-05-16&21 \\
Collections&a270ff6&2015-06-04&26415&2764   &23781&9266& 33047    &2015-09-28&3 \\

Math&41ba9e0&2006-06-05&9479&1074   &17940&14824&  32764   &2006-07-06&105 \\
Compress&004124a&2009-03-26&6741&105    &14004 &14796& 28800  &2009-03-30&46\\
Gson&c6a4f55&2010-11-02&5418&992   &8458 &8887 &  17345  &2015-10-22&17\\
JacksonXml&2d7683e&2016-01-06 &4683&436  &8012&5442& 13454  &2016-06-09&5 \\
Codec&52d82d1&2008-04-27&2584&258   &4283&8202&  12485  &2009-07-13&17\\
Jsoup&27a52f9&2011-07-02&2546&146 &3829&3151&6980 &2011-07-02&92 \\

Mockito&c1f2c4e&2009-07-09&5506&1060 &4190 &2283 &  6473 &2009-11-08&37\\
Cli&b0e1b80&2007-05-15&1937&152  &2961&3148& 6109 &2007-05-22&38 \\

Csv&de1838e&2012-03-27&806&79  &1210&1393&2603&2013-04-08&15\\
\hline
Total&-&-&327194&25812&631015&408858&1039873&-&818\\
\hline

\hline
\end{tabular}

\caption{Training and testing datasets used in our experiments. }	
\label{tab:dataset}

\end{table*}

\subsection{Diagnostic Extraction }
\label{sec-diagnostic}
\approach generates training samples that are guaranteed to be buggy (recall Section \ref{sec-perturbation}). 
For each perturbed program $\mathcal{P'}$, \approach executes it against the compiler and available test suite to determine whether it is a valid buggy training sample. 
Next, \approach extracts a diagnostic $\mathcal{D}$ of the error.
The diagnostic may be a compiler error diagnostic ($D_{c}$) or a test failure diagnostic ($D_{t}$). 

If a perturbed program does not compile, $D_{c}$ is the compiler error message.
If a perturbed program $\mathcal{P'}$ is compilable, then $\mathcal{P'}$ is executed against the available test suite. 
If one test case fails, this results in a test execution diagnostic ($D_{t}$).

%% goal of the paragraph? novelty claim

This diagnostic extraction is a key novelty of our work. 
The intuition is that they provide signals related to the bug type. The diagnostics enable the neural patch generator to generate patches according to tokens of the error diagnostic. 
No previous supervised learning-based repair approaches \cite{Tufano-ICSE19,Tufano-tse19,SEQUENCER,CoCoNuT,CURE-icse21,Recoder} include diagnostics into the input representation.

% how we do it
\approach represents the diagnostics by concatenating them with the context and buggy code (see the input representation in  Section \ref{sec-overview}).
This means that diagnostics is a considered token sequence, tokenized the same way as the code to leverage the reference to code elements and literals in the diagnostic.
The token sequence is separated into four parts.
The first part is a special token (\texttt{[CE]} for compiler errors and \texttt{[FE]} for test execution errors).
The second part is the error type, which means in our case the type of the thrown exception: runtime exceptions (e.g.,  \texttt{NullPointerException})  or test-driven exceptions (e.g., \texttt{AssertionFailedError}).
The third part is the error message and the last part is the failing test method name.
An example of a test execution diagnostic in Java would be as follows:

\[
\tiny
   \overbrace{
    \underbrace{[FE]}_\text{Special token} 
    \underbrace{ComparisonFailure}_\text{error type} 
    \underbrace{expected:1 but was:0 }_\text{error message}
     \underbrace{testEquals}_\text{failing test method name}
   }^\text{Example of Test Execution Diagnostic ($D_{t}$)}
 \]

%%%%%
Finally, we note that compiler errors are not directly related to our final goal of repairing functional errors (see ``Inference Phase'' in Section \ref{sec-overview}).
However, we have strong conceptual arguments and empirical evidence for doing so.
The compiler errors are useful for providing training samples related to project-specific typing and related to scoping information.
For example, typical compiler error diagnostics obtained from perturbation are \texttt{cannot find symbol} and \texttt{incompatible types}. They help the neural network to capture the fact that some identifiers cannot be used in a certain context.

%====================Input Representation==============
%======================================================
% \subsection{Tokenization}

% today:
% 1) naive pre-tokenization with replace
% 2 sentencePiece

% pre-tokenization is likely (very) important to remove formatting noise

% \todo{replace naive pre-tokenization by proper tokenization there are different Java tokenizers to be reused}

\subsection{Neural Architecture and Training}

A training sample consists of an expected output $\mathcal{P}$, i.e., the correct program without being perturbed (see Section  \ref{sec-overview}), 
a perturbation-based program $\mathcal{P'}$ (see \autoref{sec-perturbation}), and an error diagnostic  $\mathcal{D}$ (see Section \ref{sec-diagnostic}). The architecture provides guarantees that all training samples have been executed at least once.

%transformer
\emph{Training.} 
\approach uses a transformer neural network, which is considered  state-of-the-art \cite{CURE-icse21, RewardRepair-icse22,CoCoNuT,modit-ase21,plbart,codebert}.
The transformer model learns a conditional probability distribution  during the training process to translate from the perturbation-based program $\mathcal{P'}$ to the expected correct program  $\mathcal{P}$. 
Given the model parameters $\theta$, the training loop aims at updating $\theta$ to maximize the probability ($\Phi$) of generating the correct code given $\mathcal{P'}$ and $\mathcal{D}$:
\begin{equation}
\footnotesize
\underset{\theta}{max}    { \Phi(\mathcal{P} | \mathcal{P'},\mathcal{D})} 
\end{equation}

\subsection{Patch Ranking}
We follow the typical neural program repair process and employ beam search for patch ranking.
The beam search is a greedy algorithm that computes the most
likely tokens and ranks the outputs by
the maximum likelihood estimation (MLE) score of the overall prediction.
Thus, \approach outputs
the ordered top $K$ most likely sequences based on the likelihood of each sequence, where $K$ is configured as beam width.

\subsection{Implementation}
We implement our perturbation model based on Spoon\cite{spoon} which consists of 85 functions and more than 5K lines of code. 
We implement \approach's patch generator with state-of-the-art Transformer based architecture \cite{T5} from HuggingFace. 
The encoder and decoder consist of 6 layers.
We configure \approach to take a maximum of $384$ input tokens from buggy and context code and generate a patch with a maximum of $76$ tokens. 
\approach is trained for 10 epochs, using a batch size of $32$ and a vocabulary size of \numprint{32128}.

\section{Experimental Methodology}

%%%%%%%%%%%%%%%%%%%%%%%%%%%%%%%%%%%%
%%%%%%%%%%%%%%%%%%%%%%%%%%%%%%%%%%%%
% \addtocounter{footnote}{0}
% \footnotetext{Same date but later commit than training commit. \todo{remove}}
%%%%%%%%%%%%%%%%%%%%%%%%%%%%%%%%%%%%

% In this section, we describe our methodology for evaluating \approach by defining three research questions.

\subsection{Research Questions}

\begin{itemize}

\item \textbf{RQ1 (Effectiveness of Self-Supervision)}: To what extent does self-supervised training compare to the state-of-the-art in program repair?

\item \textbf{RQ2 (Project-specific Training)}: To what extent does project-specific training contribute to the overall effectiveness? 

\item \textbf{RQ3 (Ablation Study)}: To what extent does each component in \approach contribute to the final effectiveness?

\end{itemize}

% ==================================
% ==================================
% ==================================
% \begin{figure}[h]
%      \begin{subfigure}[b]{0.46\textwidth}
%         \centering
%          \includegraphics[width=\textwidth]{Figures/ce.pdf}
%          \caption{The compiler error diagnostic distribution.}
%          \label{fig:ce-analysis}
%      \end{subfigure}
%      \hfill
%      \begin{subfigure}[b]{0.46\textwidth}
%          \centering
%          \includegraphics[width=\textwidth]{Figures/fe.pdf}
%          \caption{The test execution diagnostic distribution.}
%          \label{fig:fe-analysis}
%      \end{subfigure}
% \caption{Errors in the generated training samples.}
% \label{figure-error-distribution}
% \end{figure}

\subsection{Training and Testing Datasets}
\label{sec:methodology:dataset}

We construct a dataset of programs $\mathcal{P}$ to seed the self-supervision loop.
Recall that \approach evaluates the perturbation-based training samples with compiler and test suite. This forces the correct programs $\mathcal{P}$ to meet the following two requirements: 
1)  $\mathcal{P}$ needs to be buildable in order to capture the compiler error diagnostics.
2)  $\mathcal{P}$ needs to have a test suite, that can be automatically executed using a test driver, in order to capture test execution failure diagnostic.

% choose d4j
Consequently, we look for benchmarks that comply with this hard compilation and testing constraints. We chose to use the widely accepted Defects4J \cite{defects4j} benchmark version 2.0, which is composed of 835 real-world buggy programs from 17 open-source projects, each of which is compilable and executable.

% highlight difference
To collect the project-specific training samples, we perturb on the correct past version of the same 17 open-source projects.
We split them into training and testing datasets by time as follows:
the training dataset of correct programs is made of the fixed version from the earliest commits by the project. 
All the remaining bugs in the same project from a later commit are used for testing. 
This is different from all the previous works on neural program repair which consider Defects4J as the only testing dataset, we leverage the bugs from Defects4J for both training and testing.

\emph{Training sample generation.} 
\autoref{tab:dataset} shows the details of our training and testing sets. %describle column
The first column gives the name of the open-source project, and the second to the eighth columns give the details of the training data including the number of perturbation-based samples generated. The last two columns give information about the testing set, including starting date of the bugs and the number of bugs. 
%row example
For example, the first row shows that for the Closure project, we use the source code in commit \texttt{6d374c3} from \texttt{2010-01-08} for generating perturbation-based training samples, which is composed of \numprint{60875} lines of source code over \numprint{5280} test cases. From this data, \approach's perturbation model generates \numprint{234244} training samples, where \numprint{142420} are training samples trigger compiler errors (CE) and  \numprint{91824} are training samples trigger functional errors (FE). 
We note that the number of training samples samples has a positive correlation with both number of source code (LOC) and test functions. This is explainable, because the source code provides the statements being perturbed and test functions specifies functional bugs. 

\begin{figure}[t!]       \includegraphics[width=0.32\textwidth,height=2.68cm]{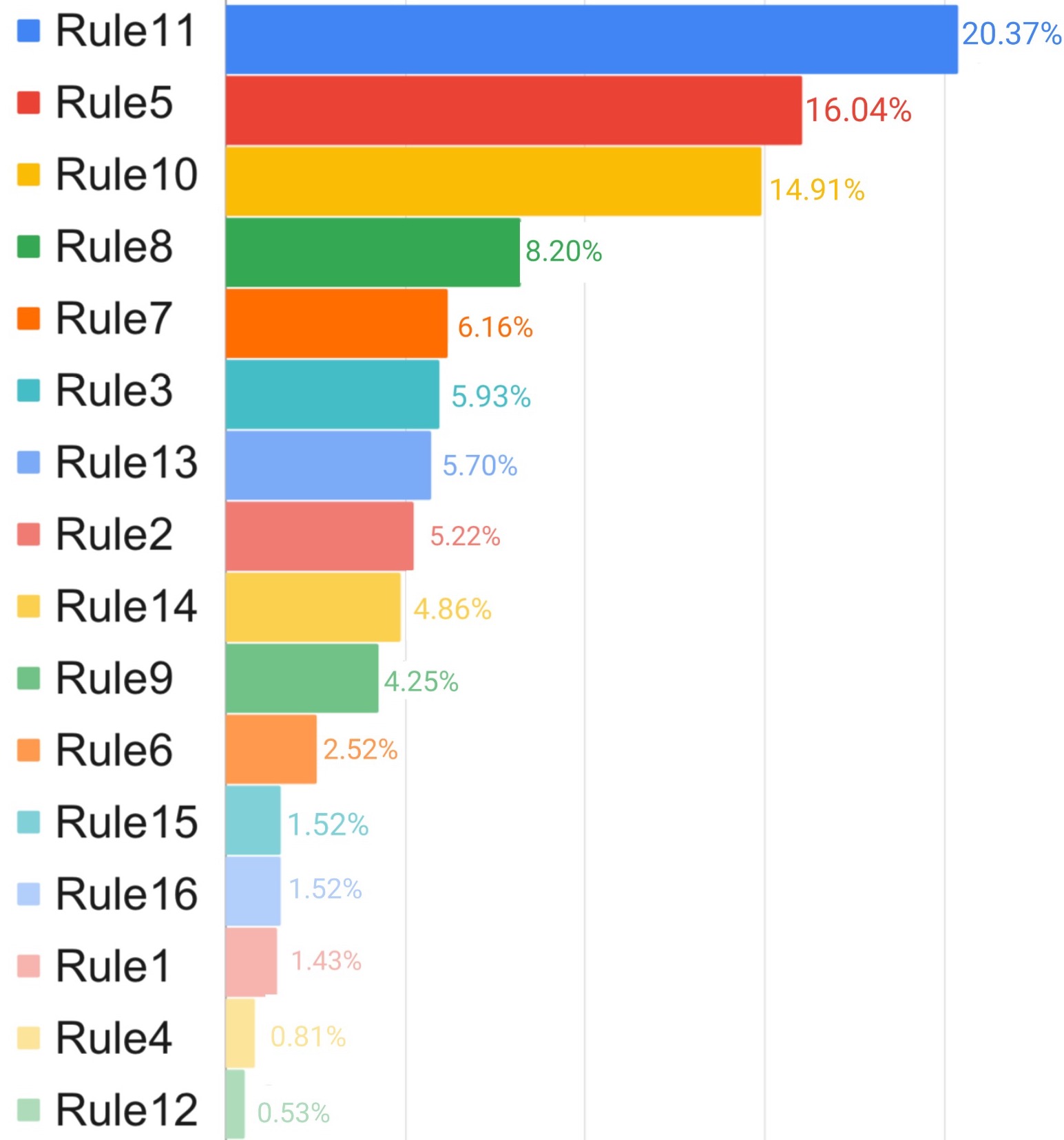}
         \caption{Distribution of training samples generated by different perturbation rules.}
         \label{fig:rule-distribution}
\end{figure}

\begin{figure}[t!]
         \includegraphics[width=0.3\textwidth,height=2.68cm]{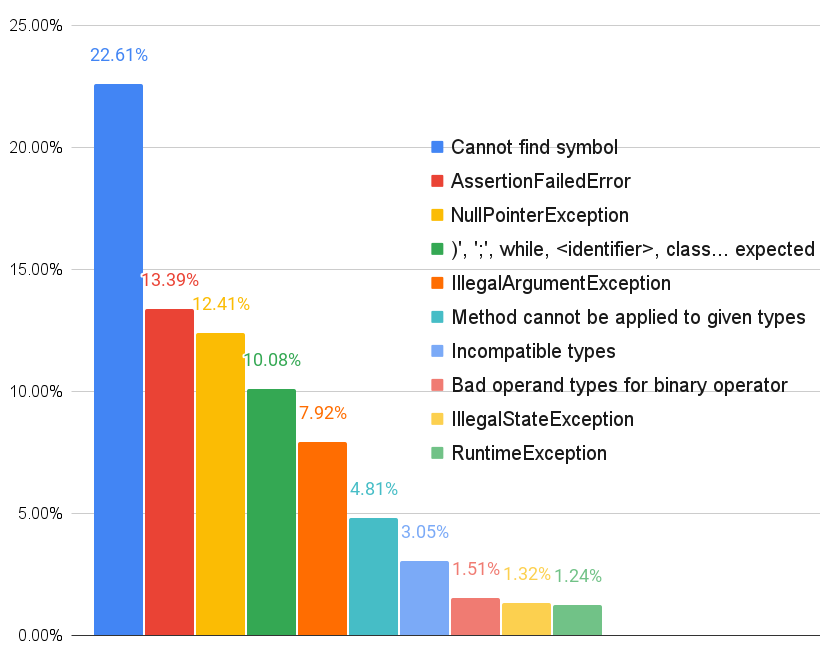}
         \caption{Distribution of errors in the generated training samples.}
         \label{figure-error-distribution}
\end{figure}

%summary training
As summarized in the last row,  we obtain \numprint{1039873} training samples, including  \numprint{631015} compiler error training samples, \numprint{408858}  functional error training samples. 
All those training samples are obtained by perturbing 17 open-source projects with \numprint{327194} lines of source code and specified with \numprint{25812} test cases in total. 
% By default, the similarity threshold is configured as \texttt{0.5}  based on Euclidean distance when transplanting related donor statements for $Rule_{6}$, $Rule_{12}$ and $Rule_{13}$. This threshold is increased to  \texttt{0.7} for $Rule_{9}$ to search similar donor statements, whose learning objective is to replace with similar statement.
Notably, the testing set is composed of all bugs from Defects4J version 2.0 minus the bugs used for training. This gives 818 (835 - 17) testing samples.

\emph{Perturbation analysis.}
\autoref{fig:rule-distribution} shows the distribution of the perturbation-based samples generated with \approach's perturbation rules. From this figure, we make the following observations:
1) All perturbation rules contribute to generating training samples  and  result in a diversity of training samples.
2) Not all the rules equally generate samples. This is explained by the characteristics of the code under the perturbation. For example, the variable perturbation $Rule_{11}$ leads to the most training samples while the  perturbation $Rule_{12}$ yields the least (wrap the target statements with a block).

\begin{table*}[t!]
\footnotesize
\renewcommand{\arraystretch}{1}

\begin{tabular}{p{0.28\linewidth}p{0.1\linewidth}p{0.07\linewidth}llllll}
\hline
 &&&  \multicolumn{2}{c}{Spectrum-based FL} &\multicolumn{2}{c}{Perfect FL }  \\

 \textbf{Approaches} & \# Training  & \# Beam  &  D4J (v1.2) & D4J (v2.0) &  D4J (v1.2) & D4J (v2.0)   \\

% &&&388 Bugs & 430 Bugs\\

\hline
Nopol \cite{nopol} (no learning) & - & - & 1/30 & - & 2/8  & -\\
DynaMoth \cite{DynaMoth} (no learning) & - & - & 1/22 & - & 3/13  & -\\

GenProg-A \cite{Yuan2017ARJAAR} (no learning) & - & - & 2/30 &  - & 12/36  & -\\
SimFix \cite{Simfix:2018} (no learning)  &-&-& 25/68 & 2/25 & 29/50\\
TBar \cite{tbar}  (no learning) & - & - & 24/72 &  8/50 & 52/85 \\
SequenceR \cite{SEQUENCER} (supervised learning) & \numprint{35578} & 50 &-&-  &14/19& -   \\
CoCoNuT \cite{CoCoNuT} (supervised learning) & \numprint{24471491} & 1000 & - & - &43/85& -   \\
CURE \cite{CURE-icse21} (supervised learning) & \numprint{24471491}   &  1000 & -&-  &55/102 &- \\
RewardRepair \cite{RewardRepair-icse22} (supervised learning)& \numprint{2307241}   & 200  & 27/- &  24/-  & 44/- & 43/- \\
Recoder \cite{Recoder} (supervised learning)&  \numprint{103585}  & 100    &\textbf{49/90} & 19/46  &64/-& -  \\
BugLab \cite{allamanis2021self-buglab} (self-supervised learning) & \numprint{415687} &50& -& - & 17/27 & 6/11 & \\
\hline
\approach  (self-supervised learning)&  \numprint{1039873}   & 50  &39/65 & \textbf{28/42}& \textbf{65/79} &\textbf{45/51} \\
\hline
\end{tabular}

\caption{ \approach's effectiveness w.r.t. the state-of-the-art over two testing datasets. In the cells, x/y : x denotes the number of correct patches, and y denotes the number of plausible patches that pass all human-written test-suite. A ‘-’ indicates that the APR approach has not been evaluated on the considered benchmark.
}
\label{tab:comparison-sota}

\end{table*}

% \addtocounter{footnote}{0}
% \footnotetext{This number is different from the number reported in original paper, because the replication package contains 2 duplicated patches (Closure-62 vs. Closure-63, and  Closure-92 vs Closure-93) and 5 patches reported as plausible and not as correct in the GitHub repository.}
%recorder closure-21 not correct

% some data about the generated errors
\emph{Diagnostic analysis.}
We now look at the composition of the diagnostics of the perturbation-based training samples.
\autoref{figure-error-distribution} demonstrates the Top-10 distribution by error type. 
It can be seen that our perturbation algorithm creates a diverse set of compiler diagnostics and functional diagnostics, with no error type dominating the other. 
% top 1 and its meaning 
The top-1 diagnostic is \texttt{cannot find symbol}, which is a semantic error by the compiler, those training samples are an important signal to the neural network to generate code with tokens complying with the available variables, types and methods according to scoping and typing constraints.
% more on functional errors.
The next two are the most common functional error diagnostics (\texttt{AssertionFailedError} and \texttt{NullPointerException}), which  show that our perturbations relate to behavior.

\emph{Dataset filtering.} Note that our testing Defects4J bugs are all caused by function errors. To ensure the close distribution between training data and testing data,  we conduct training dataset filtering as follows: 1) keep the \texttt{[FE]} samples and those \texttt{[CE]} samples with semantic errors, and 2) remove \texttt{[CE]} samples with syntactic compilation errors, e.g., ``; is expected ''.

\subsection{Patch Verification}

The patch verification follows the existing related work  \cite{SEQUENCER,CoCoNuT,CURE-icse21,Recoder}. We first execute all patches with the compiler and human-written test suite to identify plausible patches.
Then, the plausible patches are executed against independent automatically generated test cases by prior work \cite{drr}.
Lastly, we manually analyze the patches based on the ground truth developer's patch. 
All manual analysis results are confirmed by two authors, to avoid human errors and author bias.
To sum up, a patch is deemed correct if 1) it is plausible according to the developer-written and the augmented test suite  \cite{drr},  and  2) it is identical to the developer patch or if it is considered as correct by manual analysis done by  two authors.

\subsection{Methodology for  RQ1 }
\label{sec:method_rq1}
In RQ1, we compare 
\approach against the state-of-the-art of 1) supervised learning repair approaches: SequenceR \cite{SEQUENCER}, CoCoNuT~\cite{CoCoNuT}, CURE \cite{CURE-icse21}, Recoder \cite{Recoder}, and RewardRepair \cite{RewardRepair-icse22}. 
\ASECMR{We do not include TFix \cite{tfix} because it is trained on JavaScript, which cannot be used to evaluate Defects4J bugs in Java.}
2) Semantics-based repair approaches: Nopol \cite{nopol} and DynaMoth \cite{DynaMoth}; 3) Search-based repair approaches: GenProg-A \cite{Yuan2017ARJAAR} (the Java implementation of Genprog \cite{LeGoues2012GenProg}), SimFix \cite{Simfix:2018} and TBar \cite{tbar}.

Moreover, we re-implement the Java version of the  self-supervised learning approach of BugLab \cite{allamanis2021self-buglab} (originally designed for Python) with all four perturbation rules regarding operators, variables and literals. Recall that the goal of BugLab is not to obtain project-specific training samples, thus it is not explicitly executed on a past version of the project under repair. For a fair comparison, we run BugLab on the same considered past projects as \approach, which results in \numprint{415687} training samples  generated.
Notably, there is no diagnostic included in BugLab's perturbation-based training samples by its construction.

We report the quantitative results from the corresponding papers and repositories \cite{Liu2020Efficiency}. Recall that we use 17 Defects4J bugs for training. For a fair comparison, we also remove those bugs from their reported results. 
We follow the related work by employing spectrum-based fault localization \cite{yang2017empirical,yang2021evaluating} (FL) Gzoltar \cite{GZoltar} and also assuming the fault has been correctly localized \cite{SEQUENCER,CoCoNuT,CURE-icse21}, an evaluation technique known as the perfect fault localization assumption \cite{Liu2020Efficiency}, so that our work could be fairly compared with theirs.
For a fair comparison, we follow the related work to use a beam search size of 50, which is the common range of considered beam width \cite{SEQUENCER,Tufano-tse19,Recoder}. We follow the related work \cite{CoCoNuT,CURE-icse21} to use ensemble training models from 10 training epochs for patch generation.
% %talk about beam
% For a fair comparison, we configure the beam search value to 100 by following related work.
% , which controls the model effectiveness separated from the effect of search space since the bigger beam search is known to produce more correct patches\cite{Tufano-ICSE19}.

We compute the two APR performance metrics on the testing dataset:
1)  the number of bugs that are correctly repaired, and
% \item the number of bugs that can be uniquely repaired by individual repair approaches.
2) the ranking information of correct patches configured by beam width in the beam search algorithm per the developer acceptance perspective shown by Noller et al. \cite{nollertrust-icse22}.

\subsection{Methodologies for RQ2 \& RQ3}
\label{sec:methodology:rq2}

In  RQ2, we evaluate the effectiveness of \approach with and without training samples from the project under repair.
Recall that in RQ1, our training set and testing set contain the same projects, while the testing test comes from the latter commits to guarantee the fairness and the practical usage of \approach (i.e., make sure the fixes are not used during the training).
Nevertheless, the training samples from a past version contain project-specific context, e.g., code tokens and similar expressions.  

Consequently, in RQ2, we exclude project-specific training samples. We create one new training set per project by discarding  perturbation-based samples from the project under repair.
For example, to test \approach's effectiveness on project \texttt{Chart}, we create a training set without the \numprint{225953} training  samples from \texttt{Chart}.
We compute the  metric about the number of correctly repaired bugs as in RQ1.
%  all six open source projects from D4J (v1.2) with

% \todo{Just one idea:It would be interesting to see how it works by training from only the project under repair (the older version)? }

In RQ3, we conduct an ablation study to evaluate the contribution of each component to \approach, specifically, we respectively remove three components from \approach:  diagnostics $\mathcal{D}$, FE training samples, and CE training samples. We train three models without each of the above components (ablation) and evaluate them on Defects4J (v1.2) per the number of correctly repaired bugs.

\section{Experimental Results}
\subsection{Answers to RQ1: Effectiveness of Self-supervision}

In RQ1, we compare the effectiveness of \approach with the state-of-the-art in APR.
\autoref{tab:comparison-sota}
shows the patch generation results of \approach and related work on  two versions of Defects4J  benchmark \cite{defects4j}: D4J (v1.2) and D4J (v2.0) with both spectrum-based fault localization (FL) and the perfect  fault localization assumption.
%coulmn 
The first column is the approach name and its bibliographic reference.
The second and the third column give the experimental setup, incl. the number of training samples and the beam search configuration for learning-based approaches. 
%Explain the table
The fourth to seven columns show the number of correct patches and plausible patches by each APR approach on the two considered benchmarks, denoted in the format of x/y.
The results are those reported in the literature, either in the original paper or in subsequent comparative experiments \cite{Liu2020Efficiency}. 
As shown, approaches repair a smaller number of bugs on D4J (v2.0) than D4J (v1.2), suggesting that  bugs from D4J (v2.0) are  more difficult to repair.

%\approach outperforms all APR approaches on both benchmarks.  
In total, \approach correctly repairs 39 bugs from D4J (v1.2) and 28 bugs from D4J (v2.0) with spectrum-based FL. This number is increased to 65 and 45 respectively when providing perfectly localized buggy locations. 
% We note that the number of correct patches is largely increased with perfect fault localization provided, yet the impact is modest in terms of the number of plausible patches. This is in line with the finding from prior work \cite{Liu2020Efficiency}.
%
Overall, our experimental results validate the novel concept of generating perturbation-based training samples based on a past version of projects under repair.
\approach outperforms all related work but Recoder on benchmark D4J (v1.2) with spectrum-based FL. This could be explained by effectiveness from different FLs and potential benchmark overfitting to D4J (v1.2) \cite{Durieux:2019:RepairThemAll}.
In the following, our comparison fully focuses on the 110 bugs repaired with perfect FL, for a fair comparison with the closely related work.

\begin{figure}[t!]         \includegraphics[width=0.35\textwidth,height=3.08cm]{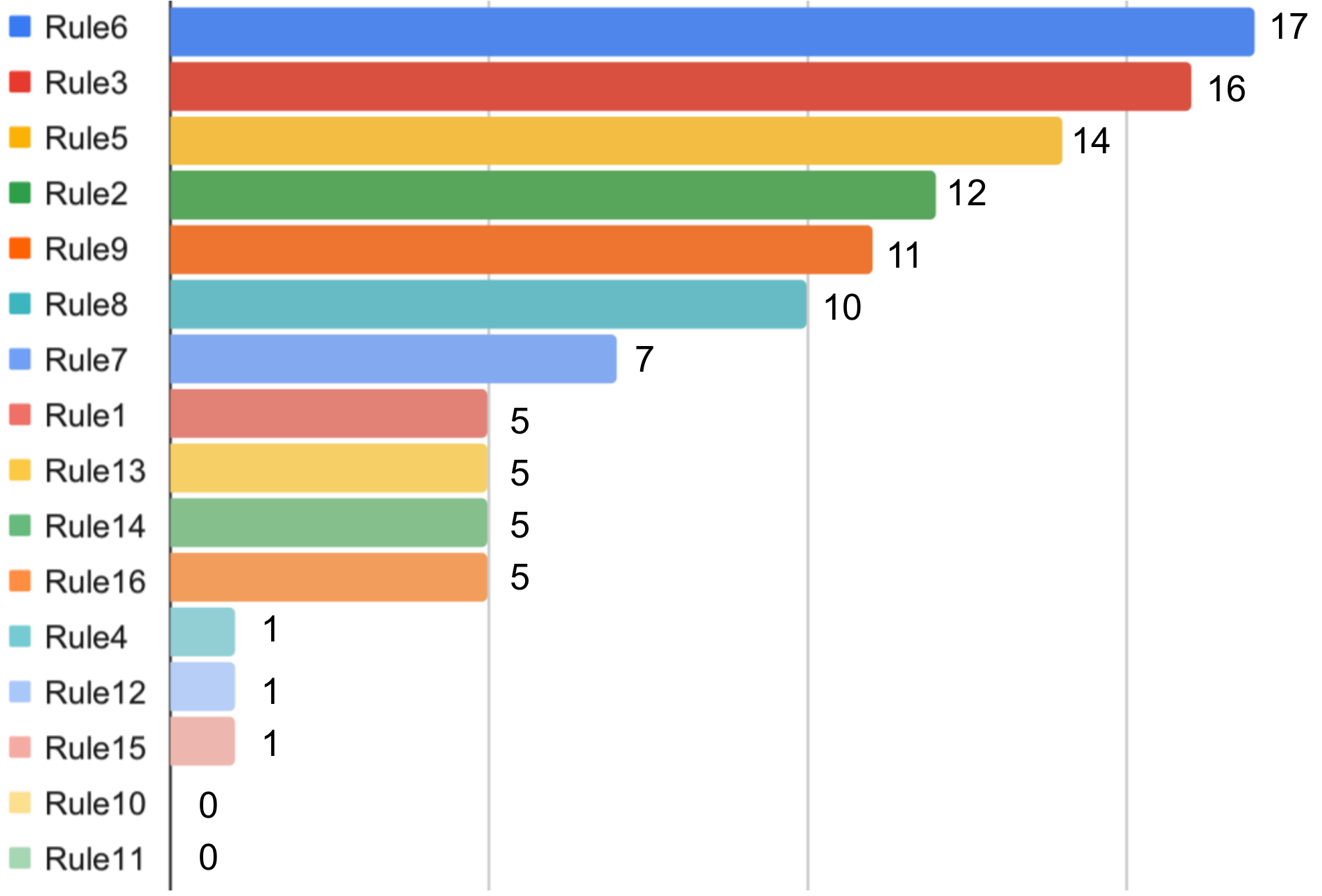}
         \caption{Correlation between the number of repaired bugs according to the corresponding perturbation rule.}
         \label{fig:correlation}
\end{figure}

%distribution.
\textbf{Correlations between repaired bugs and perturbation rules.} We look at  how the 110 repaired bugs correlate to the perturbation rules.
We map the repaired bugs to the corresponding perturbation rules based on manual analysis.
\autoref{fig:correlation} gives the result as a bar chart.
The 14/16 perturbation rules contribute to repair at least one buggy program, showing their usefulness and complementarity.
We note that the number of training samples for a rule is not linearly related to the corresponding bug.
For example, the top-1 repaired bug type relates to $Rule_{6}$: restrict/relax wrong boolean expressions, however, the proportion of training samples generated by $Rule_{6}$ is low compared to the rest.
This suggests that learning happens across perturbation rules. 
% Correlated to our four learning objectives (see Section  \ref{sec-perturbation}),   \autoref{fig:correlation} verifies \approach learns to: 
% 1) use valid identifiers to repair 65 bugs, e.g., by $Rule_{5}$;
% 2) reuse existing code to repair 28 bugs, e.g., by $Rule_{6}$;
% 3) synthesize code according to the context to repair 11 bugs, e.g., by $Rule_{16}$;
% 4) delete code to repair 6 bugs, e.g., by $Rule_{13}$.

\textbf{Comparison to BugLab.} 
BugLab \cite{allamanis2021self-buglab} is the closest related work.
As seen in \autoref{tab:comparison-sota}, \approach outperforms BugLab by a large margin.
The reason comes from the perturbation model. The one of BugLab is narrow and restricted to specific bug types.
On the contrary, \approach's perturbation model considers more rules and they are generic in nature. This results in diversity and genericity of training samples. In particular, no code transplantations and no deletions are considered in BugLab. This not only decreases the chances of learning to insert and learning to delete, but also fails to generate more training samples with project-specific knowledge (usage of domain types and methods).

%ranking.
\textbf{Deeply integrated prioritization.} 
\approach outperforms the related work on search-based and template-based repair approaches.
One key reason may relate to enumeration and prioritization. In search-based and template-based repair, one has to enumerate all possible solutions for a given transformation point or template hole. To prioritize some patches, ad hoc solutions based on heuristics are baked into the enumeration.
On the contrary, \approach has built-in prioritization.
\approach learns to prioritize  repair actions in a fully data-driven manner based on the training set, with no manually defined prioritization rules or heuristics \cite{LeGoues2012GenProg,ICSE18-patchsim}. 
This results in a natural and effective patch ranking.  \autoref{tab:ranking} shows the distribution of the 110 correct patches' ranking position output by \approach. We can see that 60.0\% of correct patches are ranked in the top-10 by the beam search algorithm.
According to recent work \cite{nollertrust-icse22}, correct patches ranked at top-10 are important for developers to accept  in practice.

\textbf{Quantity versus quality of training samples.}
\approach is trained on fewer samples than CoCoNuT, CURE and RewardRepair, yet yields better performance. This  suggests  the perturbation-based training samples are of higher quality and contain more information. There are two reasons for it:  project-specific knowledge and no noisy commits.
For supervised program repair, the past commits used for training from GitHub suffer from many limitations: they are not guaranteed to be atomic bug fix commits \cite{fse21commit} and they may include unrelated code changes (e.g., new functions, comments and optimization, etc).
On the contrary, all training samples generated by our perturbation model are controlled and guaranteed to bug-triggering samples with no unrelated changes.

\textbf{Uniquely repaired bugs.} Compared with all learning-based repair approaches, \approach uniquely repairs 10 bugs that have never been
reported as repaired by other learning-based approaches, while the other three approaches CURE, Recoder and RewardRepair also uniquely
repair respectively 6, 9, and 2 different bugs.  We manually look at the patches for those 10 uniquely repaired bugs, that are repaired thanks to the learned project-specific knowledge and explicit test diagnostics. The result shows the complementary between \approach and other supervised learning approaches, which further suggests the usage of combination training samples from self-supervised learning and supervised learning, even pre-training models \cite{natgen-fse22}.

\begin{table}[t!]
\footnotesize
\renewcommand{\arraystretch}{1.3}

 \begin{tabular}{p{0.12\linewidth}p{0.12\linewidth}p{0.12\linewidth}p{0.12\linewidth}p{0.12\linewidth}p{0.12\linewidth}}
 \hline
 Top-1 & Top-5 & Top-10 & Top-20 & Top-30 & Top-40 \\
(beam=1) & (beam=5) & (beam=10) & (beam=20) & (beam=30) & (beam=40) \\
 \hline

30.8\% & 53.8\% & 60.0\% & 75.4\% &  83.1\% &  92.3\%  \\
 \hline

\end{tabular}
\caption{The ranking information of correct patches w.r.t the beam search configuration.}
\label{tab:ranking}

\end{table}

% \textbf{Impact of FL.} 

% In learning-based repairs, context code has been shown important. Different from related work that context comes from surrounding code of buggy method, \approach sees a larger context of project, thanks to perturbing on a past version of projects.

% \textbf{Limitations.}
% \approach fails to repair the 420 multi-location bugs that require to correctly repair bugs at different locations in one repair attempt. 
% \approach learns a single objective from training samples, rather than the interactions of bugs fixing in multi-locations. 
 % In theory, self-supervised learning is powerful to learn all code. In practise, self-supervised learning could also suffer from exposure bias due to the limited code in a project under repair. 

\fbox{
\parbox{0.438\textwidth}{
Answer  to  RQ1: 
\approach correctly fixes 65 and 45 bugs for D4J (v1.2) and D4J (v2.0) respectively. This state-of-the-art performance is explained 
by 1) \approach's novel project-specific training loop, providing essential domain knowledge for repair (project-specific tokens and their semantic relationships);
2) \approach's novel input representation based on execution diagnostics.
}}

\subsection{Answers to RQ2: Project-specific Training}

In RQ2, we explore in-depth the importance of project-specific training, per the original protocol described in Section \ref{sec:methodology:rq2}.
\autoref{tab:project-specific} shows the effectiveness of \approach with and without project-specific training samples. The first column gives the test project of D4J (v1.2). The second column shows the number of bugs correctly repaired without project-specific training samples. The third column shows the number of bugs repaired by including project-specific samples, summing to 65 as reported in \autoref{tab:comparison-sota}. The improvement percentage is given in the last column.

Over all six projects, \approach's model without project-specific training samples correctly repairs 45 bugs.
On the contrary, \approach with project-specific training samples correctly repairs 65 bugs, which represents 20 more bugs and an overall improvement of 30.8\%.
This shows that project-specific training with perturbation-based training samples for the project under repair is valuable. 
The largest improvement is for project \texttt{Closure} (+40.0\%).

\begin{listing}[t!]

\begin{lstlisting} [firstnumber=419,backgroundcolor=\color{white},basicstyle=\scriptsize\ttfamily]
 
<@\colorbox{red!40!}{-  for (int i = 0; i < weights.length; i++) \{ \quad\quad\quad\quad\quad\quad\quad\quad\quad\quad\quad\quad
}@>    
<@\colorbox{green!40!}{+  for (int i = begin; i < begin + length; i++) \{ \quad\quad\quad\quad\quad\quad\quad\quad\quad\quad
}@>   
                               
                                                
                                    

\end{lstlisting}

\caption{\approach's patch for Math-41, reusing statements in project-specific samples.}
\label{lst:math41}

\end{listing}

% otherclass Skewness.java 

% buggy class = Variance.java

\begin{listing}[t!]

\begin{lstlisting} [firstnumber=419,backgroundcolor=\color{white},basicstyle=\scriptsize\ttfamily] 

<@\colorbox{red!40!}{-  if (target != null )  \{\quad\quad\quad\quad\quad\quad\quad\quad\quad\quad\quad\quad\quad\quad\quad\quad\quad\quad\quad\quad\quad\quad\quad}@>
<@\colorbox{green!40!}{+       if (target != null \&\& target.getType() == Token.STRING ) \{\quad\quad\quad\quad\quad}@>
\end{lstlisting}
\caption{Patch for Closure-57 only repaired by \approach. }
\label{lst-closure57}

\end{listing}

\begin{listing}[t!]

\begin{minipage}[b]{0.49\textwidth}
\begin{lstlisting} [firstnumber=419,backgroundcolor=\color{white},basicstyle=\scriptsize\ttfamily] 

for (int i = 0; i < array.length; i++) {            
<@\colorbox{red!40!}{- classes[i] = array[i].getClass(); \{ \quad\quad\quad\quad\quad\quad\quad\quad\quad\quad  \quad\quad\quad\quad\quad\quad }@>        
<@\colorbox{green!40!}{+ classes[i] = array[i] == null ? null : array[i].getClass();  \quad\quad\quad\quad }@>          
\end{lstlisting}

\subcaption{\approach's patch for Lang-33, identical to the human-written patch. }
\end{minipage}%
\hfill
\begin{minipage}[b]{0.49\textwidth}
\begin{lstlisting} [firstnumber=419,backgroundcolor=\color{white},basicstyle=\scriptsize\ttfamily] 
<@\colorbox{yellow!40!}{Diagnostic: [FE] java.lang.NullPointerException \quad\quad\quad\quad\quad\quad\quad\quad\quad\quad\quad}@> 
    \end{lstlisting}
\subcaption{Diagnostic for bug Lang-33}
\end{minipage}%
\caption{\approach's patch for Lang-33 guided by diagnostics.}
\label{listing-lang-33}
\end{listing}

\textbf{Project-specific training samples with reusable statements and expressions.}
\autoref{lst:math41} gives a \approach's patch for bug \texttt{Math-41}, which is identical to the human-written patch.  This bug is only repaired by including training samples
from the project under repair, here \texttt{Math}. 
Fixing this bug is non-trivial, because it requires correctly updating the initialization value of \texttt{i} from \texttt{0} to \texttt{begin}, and correctly updating \texttt{weights.length} to \texttt{begin+length}, in a single fixing attempt. 
By analyzing the project-specific training samples, we see that the same fixing for-loop statement (in green) appears over 300 times. 
Notably, the version of project \texttt{Math} used for self-supervised
training was from \texttt{2006-06-05}, and the bug \texttt{Math-41} was being fixed
on \texttt{2011-11-30}. Despite learning on a five years old version, \approach captures valuable information from the project-specific training samples and makes a valuable contribution to a new and unseen bug appearing five years after. This clearly shows that \approach succeeds in capturing project-specific knowledge in the neural network.

\textbf{Project-specific training samples enable learning semantic relationship between unique tokens.}
A past version not only enables the neural model to learn to use unique fixing tokens, it also enables the network to capture their semantic relationships.
For example,  \autoref{lst-closure57} shows a bug from \texttt{Closure-57}  only repaired by \approach in the literature. 
In this bug, there is no identical fixing expression \texttt{target.getType()==Token.STRING} in the training samples.  Yet, the two project-specific tokens \texttt{target.getType()} and \texttt{Token.STRING} separately appear in the training samples. \approach learns to repair the bug with the semantic relationship (co-occurrence) of these expressions based on unique tokens. 
This bug is only repaired by including project-specific training samples. 
\begin{table}[t!]
\footnotesize
\renewcommand{\arraystretch}{0.95}

\begin{tabular}{p{0.15\linewidth}ccr}
\hline
\multirow{2}{*}{Project}  &  \approach &   \approach  &  \multirow{2}{*}{Improvement} \\
  &  w/o Project &   with Project & \\
\hline
Chart  & 6 &  7  &   +1 (+14.3\%)\\
Closure  &  12 & 20  & +8 (+40.0\%) \\
Lang  & 7 &10  & +3 (+30.0\%) \\
Math  & 16 &22  & +6 (+27.3\%) \\
Mockito  & 2 &3 & +1 (+33.3\%) \\
Time  &   2  & 3   & +1 (+33.3\%) \\
\hline
Total & 45 &65 & +20 (+30.8\%) \\
\hline 
\end{tabular}
\caption{ Effectiveness of \approach with and without project-specific training samples.}
\label{tab:project-specific}

\end{table}
%chart miss char-9
%lang-8 lang-21 lang 
%  Math: 41 49,98

%%%% a repair repair project-specific context or not or not

\fbox{
\parbox{0.438\textwidth}{
Answer  to  RQ2:
Project-specific training on a past  version of the project under repair  contributes to the 30\% effectiveness improvement  of \approach.
This novel and original training strategy is a key:
1) for helping the model  identify rare, yet important project-specific tokens, that may be outside the buggy context at inference time, and 
2) for encouraging the model to reuse valuable domain-specific statements and expressions from the program under repair.
}}

\subsection{Answers to RQ3: Ablation Study}
\label{sec:res-rq3}
In this RQ, we do an ablation study by training \approach on training sample with only compiler errors (CE), with only functional errors (FE),  without including diagnostics (\approach w/o D).
\autoref{tab:ablation} gives the corresponding results. 
As shown in \autoref{tab:ablation}, the three ablated models respectively repair 34, 59, 56 bugs for Defects4J (v1.2), which are fewer than the whole \approach model, which repairs 65. 
This demonstrates the necessity of each and every component.

Specifically, we make the following observations:
First, the FE training samples yield higher performance, and hence have a better value than CE training samples. This is because the neural model cannot learn enough repair actions from CE samples only, for example, an operator replacement rarely causes a compiler error.
Yet, CE training samples are important, because they provide more training samples encoding project-specific knowledge, and they help to capture typing and scoping constraints that the compiler checks.
Second, having the diagnostic in the input representation is essential for \approach to generate 9 more correct patches. 
For example, the bug \texttt{Lang-33} shown in 
\autoref{listing-lang-33},  is only repaired by \approach trained with FE training samples, clearly guided by the  \texttt{NullPointerException} diagnostic.

\begin{table}[t!]
\footnotesize
\renewcommand{\arraystretch}{1}

\begin{tabular}{p{0.138\linewidth}cccr}
\hline
 \multirow{2}{*}{Project} & \approach &\approach &\approach & \multirow{2}{*}{\approach} \\
  & only CE &  only FE &  w/o D & \\
\hline
Chart & 4 (-42.9\%) & 7 (-0\%) & 7 (-0\%) & 7 \\
Closure & 10  (-50.0\%)  & 16 (-20.0\%) & 17 (-15.0\%) & 20 \\
Lang  & 6 (-40.0\%) & 9 (-10.0\%) & 8 (-20.0\%) & 10 \\
Math  & 12 (-45.5\%)& 21 (-4.55\%) & 19 (-13.6\%) & 22 \\
Mockito  & 1 (-66.7\%)  &2 (-33.3\%) &  2 (-33.3\%)  & 3 \\
Time & 1 (-66.7\%)  &4 (+33.3\%)  & 3 (-0.0\%)   & 3   \\
\hline
Total & 34 (-47.7\%) & 59 (-9.23\%) & 56 (-13.8\%) & 65 \\

\hline 
\end{tabular}
\caption{Ablation study for \approach.}
\label{tab:ablation}

\end{table}

%recorder closure-21 not correct

\fbox{
\parbox{0.438\textwidth}{
Answer  to  RQ3:
All components of \approach are important: 
the compiler error training samples, the functional error training samples, and the diagnostics.
The most important component to the final effectiveness of \approach
is the input representation with test execution diagnostics.
}}

% \section{Discussion}

% \subsection{Impact of Fault Localization}

% By executing fault localization GZoltar \cite{GZoltar} to locate suspicious buggy lines, \approach succeeds in repair 38 bugs from D4J (v1.2) and 25 bugs from D4J (v2.0). We do not consider spectrum-based localization in RQ1 because the related work \cite{CURE-icse21,CoCoNuT} were only executed by assuming perfect buggy lines, which has been shown as a fair comparison without the affect by the effectiveness of fault localization techniques \cite{Liu2020Efficiency}. 

\section{Discussion}
\label{sec-discussion}

\subsection{Threats to Validity} 

An internal threat relates to 1) our implementation of the perturbation model may contain bugs that could prevent generating more appropriate perturbation-based training samples and 2) manual patch correctness assessment. To mitigate these threats, we make the perturbation tool and generated patches publicly available for further assess with related techniques \cite{tian2022change,tian2022best,tian2022predicting,ODS,ICSE18-patchsim}.
A threat to external validity relates to whether the performance of \approach generalizes to arbitrary programming languages. Per the standards of the field, our approach has been tested in one language (Java) and the evaluation is carried out on well-established benchmarks. In principle, our approach can be applied to other programming languages and datasets.

\subsection{Multi-location Bugs} 
Repairing bugs spread in different locations (e.g., multi-hunk bugs) remains challenging for the program repair community. Our work \approach, as much as the related work does not succeed in repairing the 420/818 multi-location bugs in Defects4J. The complexity of repairing multi-location bugs not only comes from the fixing tokens and expressions, but also from the interaction between fixes in different locations. To our knowledge, only three prior work from semantics-based and search-based approaches target on multi-location bugs: Angelix \cite{Angelixicse16}, VarFix \cite{varfix} and Hercules \cite{hercules}.
% For learning-based program repairs, this is an open challenge to explore edit interactions and to find a good way of judging partial correctness. 
% Nevertheless, 
% we believe that \approach's diagnostics can be useful for judging the partial correctness of patches, which could be a valuable signal for repairing multi-location bugs.

% \subsection{Limitations}
% \ASECMR{
% We discuss two major limitations in this study. First, the correct versions we used for generating training samples are many years older than the testing programs. This leads to the training samples do not contain the fixed ingredients of the given bugs and consequently cannot generate useful perturbation-based training samples. 
% Second, another potential limitation is that \approach requires one correct history version of the given bug and test suite to collect the diagnostics of the generated training samples, which sometimes might be unavailable.  
% }

\section{Related work}
\label{sec:rw}

We have already discussed in Section \ref{sec:background} the recent related work on APR and in Section \ref{sec:method_rq1} about close related work on neural program repair with supervised training.

\subsection{Creation of Perturbed Programs}
There are other techniques and usages for perturbating programs.
For example, the mutants of mutation testing tools \cite{mutationtesting,major} can be considered as perturbed programs. However, the mutation testing operators are meant to emulate likely programmer errors. In this paper, the goal of perturbation is entirely different, it is to create valuable training data points according to specifical learning objectives. 
Patra and Pradel \cite{bugseeding} create perturbed programs to complement mutation testing with a neural approach. They use learned token embeddings that encode the semantic similarities of identifiers and literals in the perturbation process.
Different from above work, our perturbation approach is not neural, it is based on code transformation at the AST level with program analysis. While their model is restricted to low-level modifications of operators, identifiers and literals, \approach generates samples with a larger functional impact based on code transplantations and deletions.

\subsection{Self-supervised Learning on Code }
Self-supervised learning based APR has been little explored.
Loriot et al. \cite{styler} devise a self-supervised learning loop to repair formatting issues.
Yasunaga and Liang \cite{Yasunaga20DrRepair} propose self-supervised learning for repairing compilation errors. 
In both cases, the idea is to do character level perturbations (e.g., replacing or deleting punctuation). 
On the contrary, we use AST level perturbations, which are a much larger scope and are more impactful.
Only our AST level perturbation can trigger functional errors and learn to repair them. 
Allamanis et al. \cite{allamanis2021self-buglab} train a bug detection and repair model called BugLab. The key differences with our work are that they do not execute the perturbation-based programs, therefore, no test diagnostics are included into the input representation of a bug.
\approach is the  first to generate training samples in a project-specific manner with a past version of the project under repair.

A line of work considers self-supervised learning for other downstream tasks than program repair, e.g., code retrieval and code summarization  \cite{infercode-icse21,Contrastive-self-Corder,codebert,plbart}. 
The perturbation strategies employed are typically based on token masking or single token perturbation. 
None of those previous works involves a  perturbation model  at the AST level that needs to respect strict constraints
of programming languages (e.g., variable scopes) as \approach does.

% Self-Supervised Contrastive Learning for Code Retrieval and Summarization via Semantic-Preserving Transformations
% \todo{add DeepDebug: Fixing Python Bugs Using Stack Traces, Backtranslation, and Code Skeletons}

\subsection{Training based on Execution}

% \todo{"Supervised Learning over Test Executions as a Test Oracle", SACSE '21}

% \todo{Test2Vec: An Execution Trace Embedding for Test Case Prioritization}

\ASECMR{
A series of works include test case execution as input to train a neural model.
Foivos et al. \cite{Foivos-SACSE21-supervisedOracle} extract test execution traces to train a neural model to learn to distinguish runtime patterns for passing versus failing executions for a given program.
Emad et al. \cite{test2vec} propose to represent the test execution traces to a fixed-length numerical vector for neural model training.
Mesbah et al. \cite{compilation-error-fse19} extract compiler diagnostic information as an input source for repairing compilation errors but do not use execution information. 
Wang and colleagues \cite{wang2018dynamic,embed-ke-PLDI20} leverage execution trace to learn semantic aspects in program embeddings. 
These works are not about repair from diagnostics, as we do in this paper.
Chen et al. \cite{chen2018executionguided} and Gupta et al. \cite{sed-nips20} propose execution-guided synthesis. 
Those works do not execute test case diagnostics as we do hence cannot capture assertion failures. Furthermore, they do not use self-supervision and perturbation-based programs to learn the semantics of errors.
}
%On the contrary, we obtain test execution capture assertion failures, which is the explicit diagnostic for program repair task.

\section{Conclusion}
\ASECMR{
We present a novel neural program repair model called \approach, which introduces two major novelties wrt the related work:
First, it uses self-supervision with automatically  generated training samples with program perturbation. 
Secondly, it adds execution information into the input representation that captures the failing assertion of the program under repair.
Thanks to those two key features, \approach repairs 110 out of 818 bugs from Defects4J. 
Notably, 10 of them were never repaired before by the supervised learning repair approaches, demonstrating the value and power of project-specific training and  test diagnostics embedded.
}
%% elegance: the last words of the paper are the last words of the title

\begin{acks}
We thank the anonymous reviewers for the insightful feedback.
This work was supported by the Wallenberg AI, Autonomous Systems and Software Program (WASP) funded by the Knut and Alice Wallenberg Foundation.
Some experiments were performed on resources provided by the Swedish National Infrastructure for Computing.
We thank Xuechun Xu from KTH Royal Institute of Technology for the support of computing resources.
\end{acks}

\newpage

%The key idea is to train neural repair model with project-specific training samples and augment test diagnostics into code representation to guide model towards generating correct patches. 
%We have conducted an extensive empirical evaluation on 17 open-source projects.  
%Our results clearly show that it is possible to
%to use self-supervised training samples to train neural model with project-specific knowledge.

% future work

% 1) multi-location neural program repair (by doing n-wise combination of FL results)

% 2) self-refinement of patches: we iteratively give the network a patch and its diagnostic

% 3) on-demand training: for a given bug, we take the correct version just before and fine-tune with it

\bibliographystyle{ACM-Reference-Format}
\bibliography{references}

\end{document}